\documentclass[11pt,letterpaper]{article}

\usepackage{hyperref}
\hypersetup{
    colorlinks=true,
    linkcolor=blue,
    filecolor=magenta,      
    urlcolor=cyan,
}

\usepackage{fullpage}
\usepackage{parskip}
\usepackage{cite}
\usepackage{amsmath,amssymb,amsfonts}
\usepackage{mathtools}
\usepackage{textcomp}
\usepackage{algorithm}
\usepackage[noend]{algpseudocode}
\usepackage[noabbrev]{cleveref}
\usepackage{titlesec}
\usepackage{bm}
\usepackage{subfig}
\usepackage{graphicx}
\usepackage{color}
\usepackage[dvipsnames]{xcolor}
\usepackage{colortbl}
\usepackage{float}
\usepackage{bbm}
\usepackage{comment}
\usepackage{tikz}
\usepackage{hyperref}
\usepackage{comment}
\usepackage[shortlabels]{enumitem}
\usepackage{caption}
\usepackage{minitoc}
\usepackage{stackengine}
\usepackage{times}
\usepackage{authblk}
\usepackage{acronym}

\newacro{NN}{Neural Networks}
\newacro{td-INDP}{time-dependent Interdependent Network Design Problem}
\newacro{NDP}{Network Design Problem}
\newacro{MIP}{Mixed-Integer Programming}
\newacro{INDP}{Interdependent Network Design Problem}
\newacro{DNN}{Deep Neural Networks}
\newacro{ReLU}{Rectified Linear Units}
\newacro{AR}{Accuracy Radius}
\newacro{MSE}{Mean Squared Error}

\setstackEOL{\\}
\makeatletter
\def\BState{\State\hskip-\ALG@thistlm}
\makeatother

\allowdisplaybreaks

\def\BibTeX{{\rm B\kern-.05em{\sc i\kern-.025em b}\kern-.08em
    T\kern-.1667em\lower.7ex\hbox{E}\kern-.125emX}}

\makeatletter
\def\blfootnote{\xdef\@thefnmark{}\@footnotetext}
\makeatother

\title{Deep Learning-based Resource Allocation for Infrastructure Resilience}

\author[1]{Siavash Alemzadeh}
\author[2]{Hesam Talebiyan}
\author[1]{Shahriar Talebi}
\author[2]{Leonardo Due$\tilde{\text{n}}$as-Osorio}
\author[1]{Mehran Mesbahi}

\affil[1]{William E. Boeing Department of Aeronautics and Astronautics, University of Washington, Seattle, WA, USA}
\affil[2]{Department of Civil \& Environmental Engineering, Rice University, Houston, TX, USA}

\date{}

\begin{document}

\maketitle

\blfootnote{Emails of the authors are: \\ \texttt{\{alems, shahriar, mesbahi\}@uw.edu} \\ \texttt{\{hesam.talebiyan, leonardo.duenas-osorio\}@rice.edu}}


\begin{abstract}
    From an optimization point of view, resource allocation is one of the cornerstones of research for addressing limiting factors commonly arising in applications such as power outages and traffic jams.
    In this paper, we take a data-driven approach to estimate an optimal nodal restoration sequence for immediate recovery of the infrastructure networks after natural disasters such as earthquakes.
    We generate data from td-INDP, a high-fidelity simulator of optimal restoration strategies for interdependent networks, and employ deep neural networks to approximate those strategies.
    Despite the fact that the underlying problem is NP-complete, the restoration sequences obtained by our method are observed to be nearly optimal.
    In addition, by training multiple models---the so-called estimators---for a variety of resource availability levels, our proposed method balances a trade-off between resource utilization and restoration time.
    Decision-makers can use our trained models to allocate resources more efficiently after contingencies, and in turn, improve the community resilience
    Besides their predictive power, such trained estimators unravel the effect of interdependencies among different nodal functionalities in the restoration strategies.
    We showcase our methodology by the real-world interdependent infrastructure of Shelby County, TN.
\end{abstract}


\section{Introduction}
\label{sec:intro}

The impact of infrastructure networks is omnipresent in our daily life, and hence, the importance of infrastructure resilience has been highlighted by decision-makers for the society on different levels \cite{USGovernment2013, DepartmentofHomelandSecurity2011}. 
Infrastructure systems fulfill the requirements of a functional economy by providing commodities and services such as supplies of water, gas, and electricity.
Therefore, abnormal operation of these networks can directly influence health and security, resulting in a considerable financial loss.
As an example, economic losses from natural disasters have passed an average of US\$250 billion per year \cite{mcglade2019global} and only in South America, \%3 percent of the GDP (approximately US\$71 billion) is invested in infrastructure to satisfy the restoration demands \cite{kingombe2011mapping}.
In this light, the resilience of infrastructure---its ability to withstand, absorb, and bounce back after contingencies---is of importance.
In particular, efficient infrastructure restoration planning significantly contributes to the resilience by reducing direct and indirect costs of restoration and the downtime.

Inefficient resource allocation---i.e., the distribution of budget, crew, machinery, or the like---during the infrastructure restoration exacerbates the complications of societal and economic interests---say, power or water outages---after a disastrous event.
Proper resource allocation, on the other hand, not only improves the normal performance of the entire network, but also blocks the transient bottlenecks that may intervene network's functionality.
Over the past few years, such systems have been rigorously modeled within the framework of complex networks, having stimulated several theoretical as well as empirical studies on network resilience \cite{callaway2000network, rinaldi2001identifying}.
Besides, resource allocation and optimization techniques have gained more attention due to the complexity arising from underlying interdependencies \cite{Kim2008, hong2016suppressing}---i.e., whether or not the functionality of one part of the network relies on the recovery of another part.

\begin{figure}[t]
    \centering
    \includegraphics[width=0.7\columnwidth]{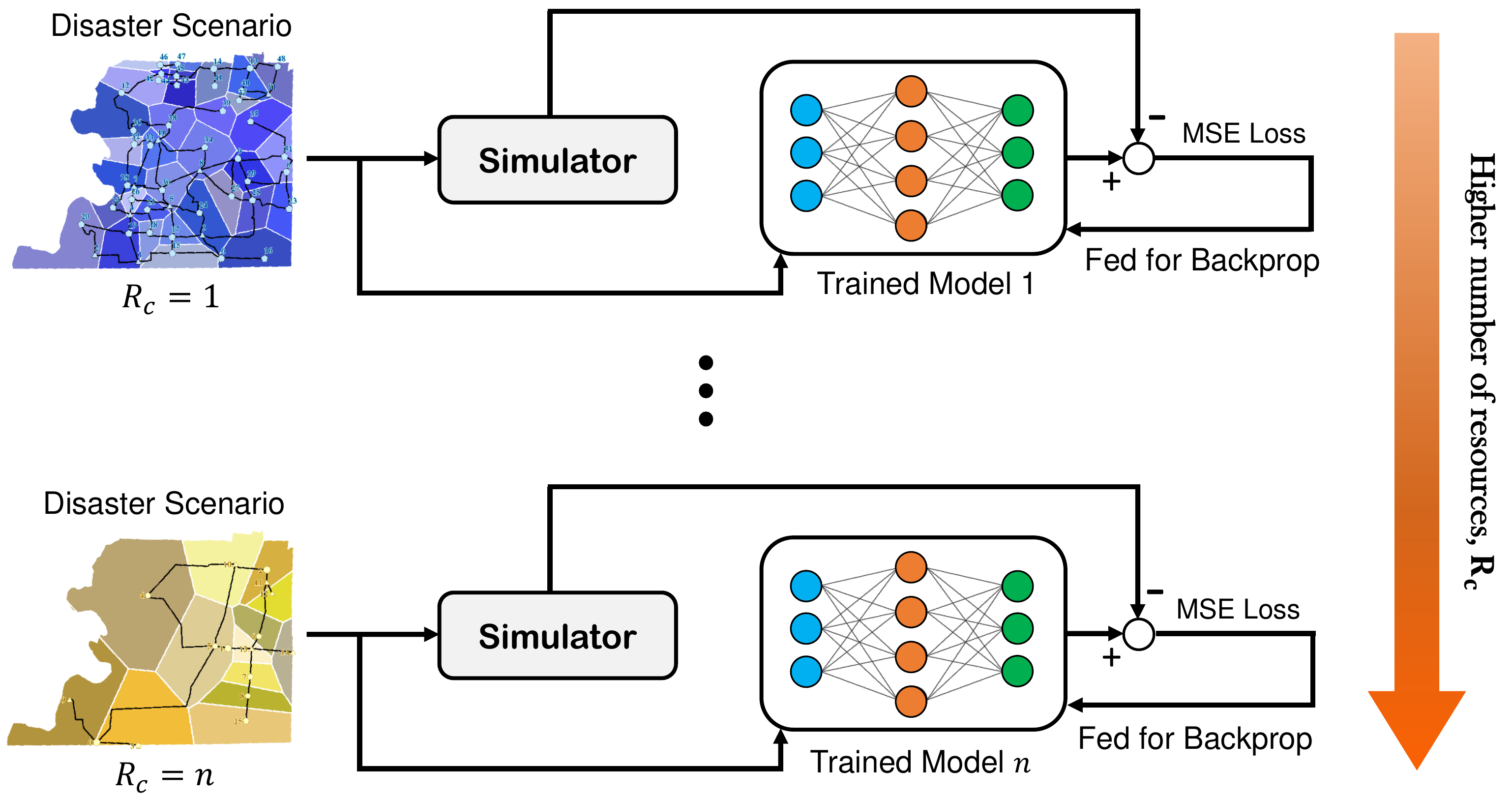}
    \caption{The schematic for multi-model training of estimators based on different damage scenarios and given number of resources, subsequently used for resilience of infrastructure.}
    \label{fig:main-model}
\end{figure}

In the meantime, providing time-dependent optimization models to determine an efficient quantity of resources and strategies may be computationally expensive, and hence, not pertinent to real-time applications \cite{Lee2007}.
Moreover, pre-calculation of prototypical disaster configurations to develop pre-disaster recovery plans becomes prohibitive as the number of configurations grows exponentially large in problem parameters.
In addition to that, fast recovery planning for time-critical applications would be costly under such computational load \cite{nurre2012restoring, gonzalez2017efficient}.
Therefore, one alternative is to create an approximation framework that would take initial disruption (damage) scenarios and their respective optimal recovery strategies as the input-output data to be leveraged in training a model that can be subsequently used to generate tailor-made recovery strategies.
In particular, we propose a deep learning-based approach to facilitate the real-time resource allocation and restoration planning of interdependent infrastructure networks.
The contributions of our work are as follows: (1) We employ \ac{NN} models and train them with high-fidelity restoration strategies devised by a \ac{MIP} formulation to predict restoration strategies in real-time, (2) we find the most efficient number of required resources based on multiple pre-trained models given the specific damage scenario (\Cref{fig:main-model}), (3) we study meaningful realizations from the learned models that provide insights about restoration dynamics and the role of network interdependencies, and (4) we provide the set of labeled data that we have collected from the simulator on Shelby County, TN testbed.
Our \ac{NN} model can then be used by decision-makers to allocate resources more efficiently in a time-sensitive situation, or when they lack fine-grained situational awareness, such as in hierarchical decision-making environments.
Also, the provided insights can guide decision-makers toward more informed long-term resilience planning.

The rest of the article is organized as follows:
We present a review of the literature in \Cref{sec:litReview}.
We provide an overview of the optimization algorithm behind the simulator and the architecture of the \ac{NN} in \Cref{sec:prelim}.
We introduce the main results along with detailed empirical analyses in \Cref{sec:results}.
Concluding remarks are then followed in \Cref{sec:conclusions}.


\section{Literature Review}
\label{sec:litReview}


\subsection{High-fidelity Restoration Model}
\label{subsec:td-INDP-model}

The restoration scenarios used to train the proposed models are devised by solving an optimization problem called the \ac{td-INDP} \cite{Gonzalez2016c}.
The solution to a \ac{td-INDP} is the least-cost time-dependent restoration strategy of disrupted interdependent networks, subject to budget, resources, and operational constraints.
The formulation of \ac{td-INDP} accounts for physical, logical, and geographical interdependencies between the networks.
Solving this problem results either in an optimal plan or a plan along with a guaranteed optimality gap.
Also, the framework accommodates networks in which multiple commodities flow as well as several levels of functionality or reconstruction of network components.

The precursor of \ac{td-INDP} is the \ac{NDP}, which has been studied since the 1960s \cite{Wong1978}.
\ac{NDP} finds a subgraph that minimizes a cost function while satisfying a set of flow and demand constraints over a single (non-interdependent) network.
NDP is an NP-complete problem \cite{Johnson1978}, which means that there is no efficient algorithm to solve it in the worst-case scenario.
Several studies extend versatile \ac{NDP}-based formulations to the setting in which several interdependent networks are involved since interdependency may significantly affect the performance \cite{Hernandez-Fajardo2011}.
In particular, here, we are concerned with studies that explore the restoration process of interdependent networks to find optimal restoration strategies.
To this end, Lee et al. \cite{Lee2007} presented an \ac{MIP} formulation that finds restoration strategy without any temporal order, and introduced a heuristic method to accelerate solving the problem.
Cavdaroglu et al. \cite{Cavdaroglu2013} employed another \ac{NDP}-like \ac{MIP} formulation for restoration and job scheduling of interdependent infrastructure systems by explicitly introducing a time index.
Gonzalez et al. \cite{Gonzalez2016} introduced \ac{INDP}, which accounts for resource constraints of the restoration process as well as the savings due to collocated restoration jobs in different networks.
To find the restoration plan for several time-steps, they solve \ac{INDP} iteratively to find the complete restoration plan.
Later, Gonzalez et al. \cite{Gonzalez2016c} introduced \ac{td-INDP}, which defines an \ac{INDP}-like problem over several (potentially all) time-steps.
Therefore, \ac{td-INDP} devises restoration strategies that consider future actions in finding the optimal strategy at the present time.
The \ac{NDP}-based models may not be applicable to time-sensitive situations due to their high computational complexity.
Nonetheless, these models can be used to study prototypical damage scenarios to identify the critical properties of the dynamics associated with the restoration process of interdependent networks.
\begin{table*}[t]
	\centering
	\caption{Parameters of the cost function used in \ac{td-INDP}.}
	\includegraphics[width=0.8\textwidth]{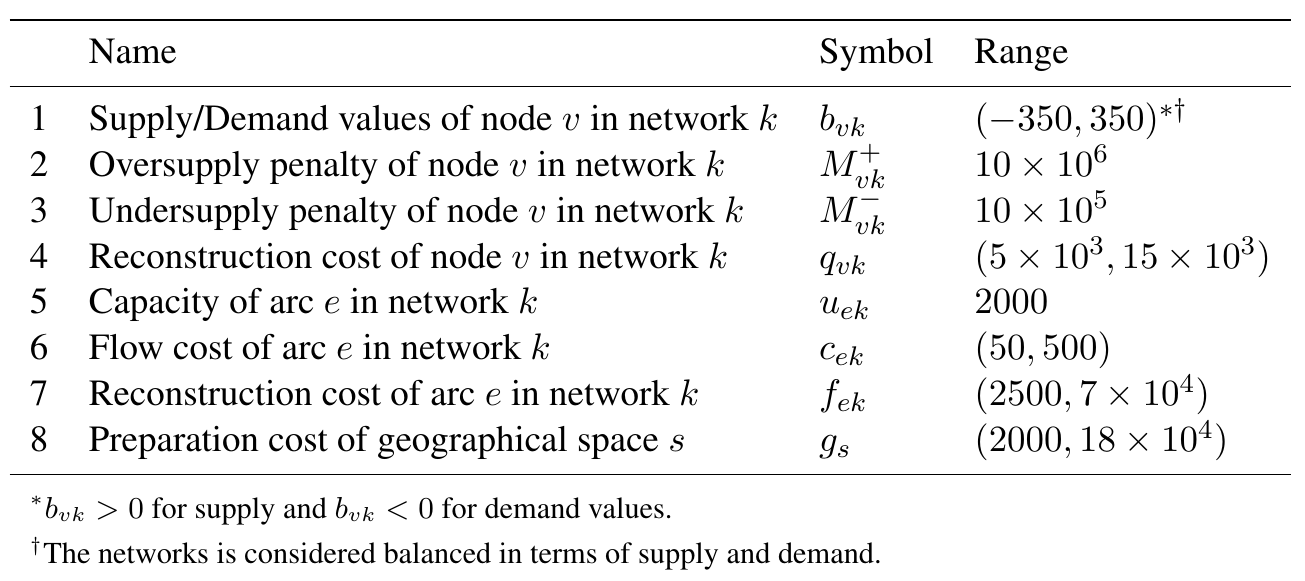}
	\label{Tab:FuncParam}
\end{table*}


\subsection{Resource Allocation}
\label{subsec:res_all_review}

Resource allocation for network restoration and recovery is usually formulated as an optimization problem.
Many of such problems are similar to the class of {\it Scheduling Problem} in the field of Operations Research \cite{Celik2016, Brucker2007}.
The class concerns scheduling the repair actions and assignment of the workforce, and includes optimization problems that are usually NP-hard \cite{Guha1999}.
In the context of infrastructure restoration, different variants of the problem have been employed, which are solved using either exact methods \cite{Guha1999,Gong2009} or heuristic and meta-heuristic approaches \cite{Fiedrich2000,Furuta2008}.
In particular, some studies \cite{Lee2007,Cavdaroglu2013,Sharkey2015c} focus on the restoration scheduling of several networks that are interdependent physically or operationally.
Scheduling problems, especially in the presence of interdependencies, usually include many intricacies that are modeled with a large set of constraints; therefore, their scope may be limited to specific applications.
In this study, we utilize td-INDP \cite{Gonzalez2016c} as our benchmark, a high-fidelity simulator which is more comprehensive while keeping the intricacies of scheduling out.
As a consequence, td-INDP has a block-matrix structure, which can be exploited to accelerate the optimization using decomposition techniques.
Moreover, we introduce data-driven models that efficiently approximate recovery scenarios from td-INDP, and hence, can be used in real-time and for time-critical situations.


\subsection{Restoration Sequence Approximation}
\label{subsec:model_prediction_review}

To proceed with the resource allocation task, we need to develop a tool for estimating \ac{td-INDP} restoration pattern.
Hence, given some specific damage scenario of a network, we use our trained models to predict the most efficient sequence of nodal recovery in the sense of optimality considered in \ac{td-INDP}.
It is well-known that such prediction (as well as structure exploitation) can be achieved by a linear operator using regression methods \cite{parsa2018hierarchical, talebi2020online, tu2014dynamic, alaeddini2018linear}.
It was shown that the linear operator embeds the interdependency of different elements in the network \cite{gonzalez2017efficient}.
Nevertheless, linear models are too simple to characterize the complex input-output behavior of \ac{td-INDP}, and thus, underestimating the majority of the spatio-temporal characteristics of the observed data.
On the other hand, deep neural networks \cite{Goodfellow-et-al-2016} capture linear and nonlinear characteristics of a given (large enough) set of labeled input/output data, making the technique suitable for a variety of applications in healthcare, science, finance, and engineering \cite{rafiq2001neural, shahid2019applications, teodorescu2008artificial, faller1996neural, wong1998neural}.
In this work, we employ \ac{NN} to address the restoration problem on synthetic data generated for different magnitudes of earthquakes and provide comparisons with the linear case.
We will further show that interdependencies among different networks (water, gas, and power) can be identified as a byproduct of our trained models with a detailed study of the \ac{NN} trained parameters.


\section{Preliminaries}
\label{sec:prelim}

\subsection{Td-INDP Algorithm}
\label{subsec:Td-INDP Algorithm}

We start introducing td-INDP with defining the interdependent network as $G\triangleq (\mathcal{V}, \mathcal{E}, \mathcal{I})$ 
where $\mathcal{V}=\bigcup\limits \mathcal{V}_k$ and $\mathcal{E}=\bigcup\limits \mathcal{E}_k$ denote sets of all nodes and arcs respectively for $k \in \mathcal{K}$, where $\mathcal{K}$ denotes the set of layers of the interdependent network and $\mathcal{I}$ is the set of interconnections.
The $k^\text{th}$ layer of the network is defined as $G_k\triangleq (\mathcal{V}_k, \mathcal{E}_k)$.
The elements of $\mathcal{E}$ represent physical connections that carry commodities while elements of $\mathcal{I}$ denote logical dependencies. 
The objective function of \ac{td-INDP} consists of four cost functions over $|\mathcal{T}|$ time-steps.
For a given time-step $t$, these functions are listed as follows: the flow cost (of commodities) $C_f^t$, the reconstruction cost of arcs and nodes $C_r^t$, the penalties due to the imbalance of supply or demand at nodes $C_u^t$, and the geographical preparation cost $C_g^t$.
The latter cost accounts for the savings from co-located restoration jobs.
All functions are linear combinations of their respective input parameter.
The mathematical form of the objective function is,
\begin{align*}
    \text{min} \hspace{2mm} \sum_{t\in \mathcal{T}}{C_f^t(c_{ek}) + C_r^t(q_{vk}, f_{ek}) + C_u^t(M^+_{vk},M^-_{vk}) +  C_g^t(g_{s})}
\end{align*}
where arguments of cost functions are introduced in Table \ref{Tab:FuncParam}, and, for our application, their ranges are adopted from the literature \cite{Gonzalez2016c,Talebiyan2020}.
The cost functions are linear, and sum up to their respective cost over all elements across the network for a given time-step $t$.
The feasible space of the solution is defined by several constraints, which can be categorized into five groups:

\begin{enumerate}[wide, labelwidth=!, labelindent=0pt]
	\item Flow conservation constraints at nodes, which guarantee the flow balance at each node.
	Their general form is,
	\begin{align*}
    	\text{Outflow - Inflow} = b_{vk} - \delta^+_{vkt} + \delta^-_{vkt},  
	\end{align*}
	$\forall v \in \mathcal{V}_k$, $\forall k \in \mathcal{K}$, and $\forall t \in \mathcal{T}$, where $\delta^+_{vkt}$ and $ \delta^-_{vkt}$ are deviation variables that record the surplus and deficit of commodities at node $v$ in network $k$ at time $t$.
	
	\vspace{1mm}
	\item Capacity constraints on commodity flow in arcs, which ensure arc flows do not exceed the capacities of arcs by using the following form $\forall e \in \mathcal{E}_k$, $\forall k \in \mathcal{K}$, and $\forall t \in \mathcal{T}$,
	\begin{align*}
    	\text{Flow} \leq \left\{
    	\begin{array}{l}
        	0 \quad \text{arc $e$ or connected nodes are non-functional} \\
        	u_{ek} \hspace{25mm} \text{otherwise}
    	\end{array}\right.
	\end{align*}
	
	\item Resource constraints, which prevent employing more resources than the Resource Cap, $R_c$, which is the number of available resources.
	In mathematical terms,
	\begin{align*}
    	\sum_{k \in \mathcal{K}} r_{kt} \leq R_c,  \hspace{12mm} \forall t \in \mathcal{T}
	\end{align*} 
	where $r_{kt}$ denotes the number of resources used for the restoration of network $k$ at time $t$. 
	A resource can represent different real-world assets such as budget, crew, or machinery. 
	In general, $r_{kt}$ is proportional to the total number of elements in network $k$ that can be repaired at time $t$ under a given restoration plan.
	In this study, we assume that repairing each element needs one resource, i.e., $r_{kt}$ is equal to the number of elements that can be repaired. 
	
	\vspace{1mm}
	\item Physical interdependency constraints, which indicate if recovery of a node depends on the functionality of others.
	
	\vspace{1mm}
	\item Co-location constraints, which ensure that any given area has to be prepared only once, even if several agents carry out reconstruction tasks inside the area.
	
	\vspace{1mm}
	\item Demand completion constraints, which ensure that certain prescribed nodes are functional only if their demands are fully satisfied.
\end{enumerate}

The explicit mathematical formulation of \ac{td-INDP} is presented in \cite{Gonzalez2016c}.
Given a set of initial states of the nodes (i.e., binary values with zero denoting damaged nodes) based on a given seismic scenario, \ac{td-INDP} determines the states of the nodes at each time-step.
For each $t\in\mathcal{T}$, the repaired nodes are those whose state is zero at $t-1$ and unity at $t$, and the number of repaired nodes is less or equal to $R_c$.
Next, we employ these spatio-temporal data---initial states and the states during the restoration process---to train our proposed \ac{NN}, which will then be used for prediction and resource allocation.


\subsection{Artificial Neural Networks}
\label{subsec:ANNs}

Given a set of labeled data, neural networks provide a computational mechanism to find a (nonlinear) mapping from a multivariate space of information to another.
We consider \ac{td-INDP} as the baseline and feed our \ac{NN} model with data generated from this simulator.
The goal is then to discover the simulator's underlying spatio-temporal behavior in order to enable an efficient, fast recovery strategy and avoid the computational complexity of repeatedly solving high-dimensional \ac{MIP}.
Our model is based on a feed-forward architecture with backpropagation used for the training.
The input layer obtains a vector of binary values that represent the functionality of the nodes, while the output layer delivers a recovery plan for the damaged nodes.
More specifically, the input layer consists of $n_i=125$ neurons representing $49$ water, $16$ gas, and $60$ power nodes in Shelby County, TN that form a $125\times 1$ vector with the value of the $j^\text{th}$ element $\nu_{ij}=0$ if the node needs restoration and $\nu_{ij}=1$ otherwise, following \ac{td-INDP} guidelines.\footnote{Our empirical observations suggest that the results will improve in case we flip these value to $\nu_{ij}=1$ for damaged and $\nu_{ij}=0$ for recovered nodes, hence, the non-damaged nodes would have computationally less effect in the output recovery plan.}
The output layer also contains $n_o=125$ neurons with the value of the $k^\text{th}$ node $\nu_{ok}\in\mathcal{T}$, denoting the time-step that the corresponding node should be recovered in the process in order to minimize the loss.
We set $\mathcal{T}=\{1,2,\cdots,20\}$ based on empirical regards.
At the prediction stage, our \ac{NN} architecture includes three hidden layers with $n_h=400$ neurons.
Later, when we explore interdependencies, we modify the settings to one hidden layer with $n_h=4$ neurons to simplify visualization.
That is, we sacrifice the prediction power of \ac{NN} as investigating one hidden layer is more intuitive.
It is straightforward to extend such interpretations to deep networks.
The error between the outcomes of \ac{NN} and \ac{td-INDP} is plugged into a \ac{MSE} loss to update the weights of \ac{NN} through backpropagation using ADAM optimizer with learning rate $\eta=0.01$.
After the training phase is completed, we perform an extensive interpretability study on the trained parameters (weights and biases) of our model.
We will discuss how such an investigation should be conducted in order to obtain useful information about the underlying behavior of the network (e.g., to infer which nodes are more influential in the recovery process).


\subsection{NN-Based Resource Allocation}
\label{subsec:NN-Based-Resource-Allocation}

Following \ac{td-INDP} settings, our goal is to find the most efficient number of resources, $R_c$, in order to get minimum-time recovery given the damage scenario (i.e., distribution of the damaged nodes, magnitudes of the earthquake, and the geographical locations).
We train a separate \ac{NN} for each possible $R_c$, from which we can address the trade-off between time-efficiency and resource utilization (see \Cref{fig:main-model}).
Subsequently, we run \ac{td-INDP} to evaluate the accuracy of the results.
Our models can then be used as an insightful reference for decision-makers to forecast and quickly adopt more efficient policies in time-critical situations.

\begin{figure}[t]
    \centering
    \includegraphics[width=0.65\columnwidth]{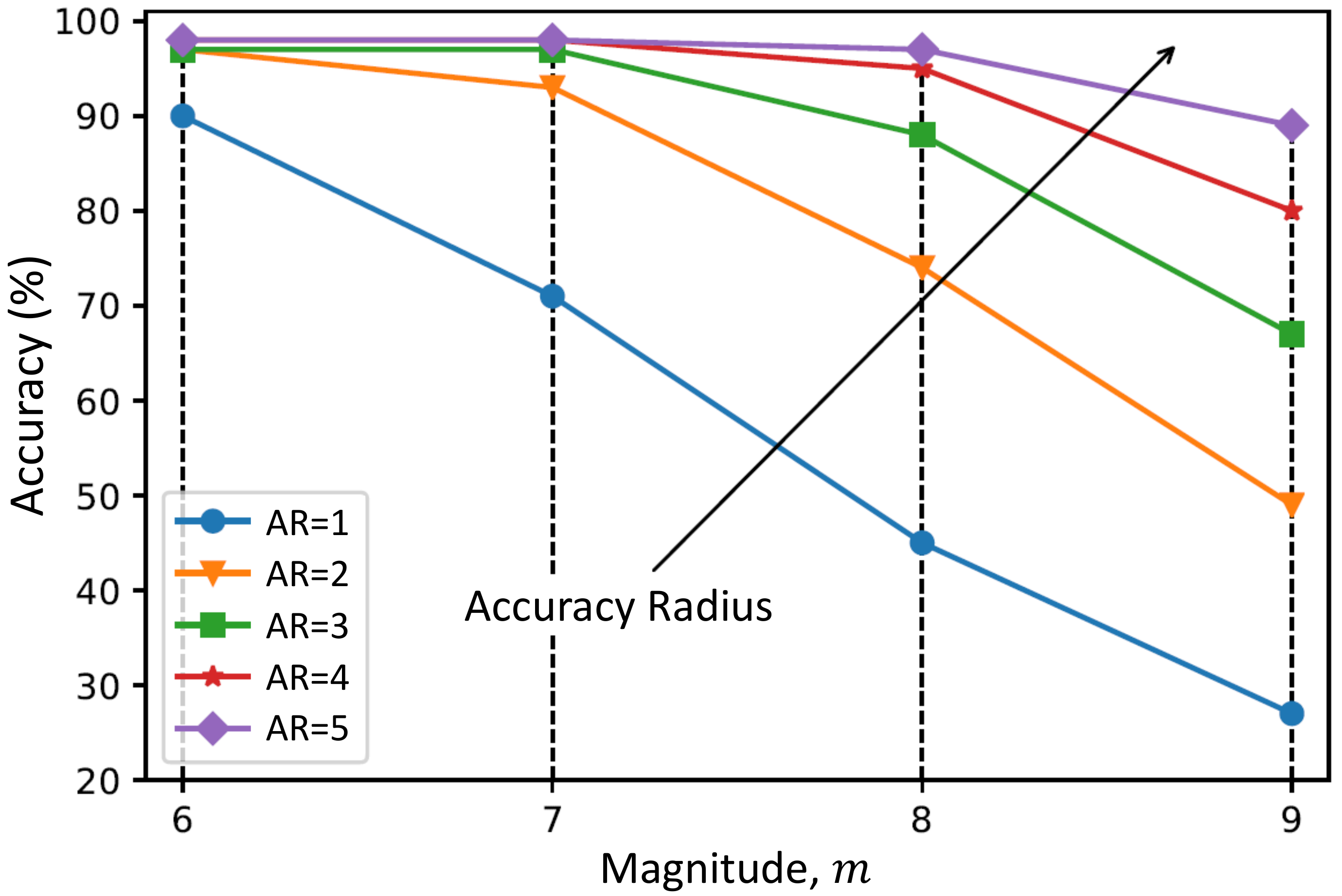}
    \caption{Accuracy vs magnitude $m$, results on the test dataset for $R_c=5$, Shelby County, TN.}
    \label{fig:accvsmag}
\end{figure}\section{Main Results}
\label{sec:results}

In this section, we provide the main results of our study.
First, we bridge between the data from the simulator (\ac{td-INDP}) and the estimator (\ac{NN} model) and provide empirical results on the accuracy of predictions.
Next, we compare these predictions for different values of $R_c$ in one plot and realize the trade-off between restoration time and the number of resources.
Finally, we discuss the interpretability of the \ac{NN} parameters.


\subsection{NN-Based Approximation of Td-INDP}
\label{subsec:NN-td-INDP}

In this part, we illustrate the qualitative results of \ac{NN} model estimates. 
As mentioned in \Cref{sec:prelim}, we feed our model with training data generated from \ac{td-INDP}.
The original batch of data contains $1,000$ damage scenarios based on simulated seismic scenarios with magnitude $m\in\{6,7,8,9\}$ ($4,000$ scenarios in total).
The seismic scenarios are computed based on the earthquake catalog of the area encompassing Shelby County \cite{Gonzalez2016}.
To make our model more generalizable, we augment the original data by perturbing the damage scenarios.
To this end, we choose a random number of nodes in the network and flip their binary values (say, we choose the $k^{\text{th}}$ node and set $\nu_{ik}\leftarrow 1-\nu_{ik}$).
Then we feed the perturbed scenarios to the simulator to obtain the labels for new training data.
Consequently, we obtain $10,000$ input-output data from the simulator for each earthquake magnitude ($40,000$ in total).
We use the augmented $40,000$ data as our training set and the original $4,000$ real scenarios as the test set.
We then evaluate the performance of our trained model by calculating an index measure called \ac{AR}, which is defined as \textit{the acceptable margin of prediction provided by the trained \ac{NN}}.
That is, for $\text{AR}=r$, the prediction of recovery time for each node is acceptable within a margin of $\pm r$ time-steps.\footnote{Designating an acceptable margin of prediction such as \ac{AR} is crucial for time-sensitive applications in which it is sufficient to acquire coarse but immediate predictions.}

The accuracy of our model prediction is plotted in \Cref{fig:accvsmag} for $R_c=5$.
As expected, the accuracy of the predictions increases as \ac{AR} gets higher.
The other unsurprising pattern---inherent to any regression-based prediction---is the monotone decline in the accuracy as the magnitude of the earthquake increments.
This behavior is inevitable as the higher intensity of an earthquake results in more damaged nodes.
This can be exponentially more challenging to learn and requires a more complex \ac{NN} structure to capture the coherency within the network.
However, for $m\leq 7$ and $\text{\ac{AR}} \geq 2$, our trained model predicts the optimal recovery sequence with more than $\%90$ accuracy.

\begin{figure}[t]
    \centering
    \begin{minipage}{.45\textwidth}
        \centering
        \includegraphics[width=0.75\columnwidth]{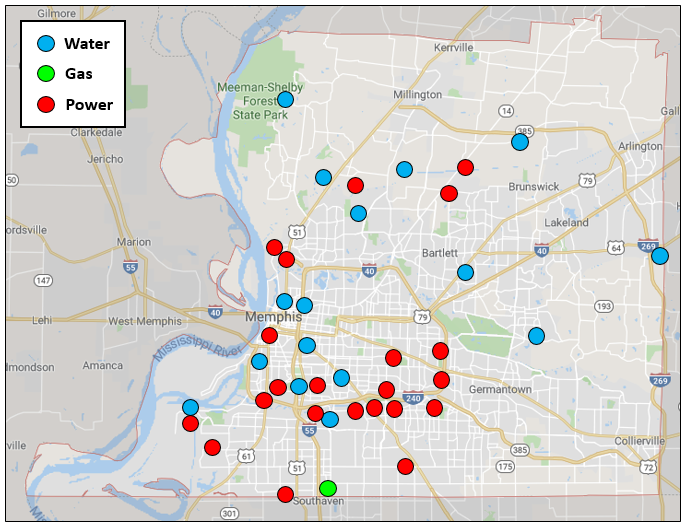}
        \caption{One sample damage scenario for $m=7$, Shelby County, TN.}
        \label{fig:damage_map}
    \end{minipage}%
    \hspace{7mm}
    \begin{minipage}{.45\textwidth}
        \centering
        \includegraphics[width=0.87\columnwidth]{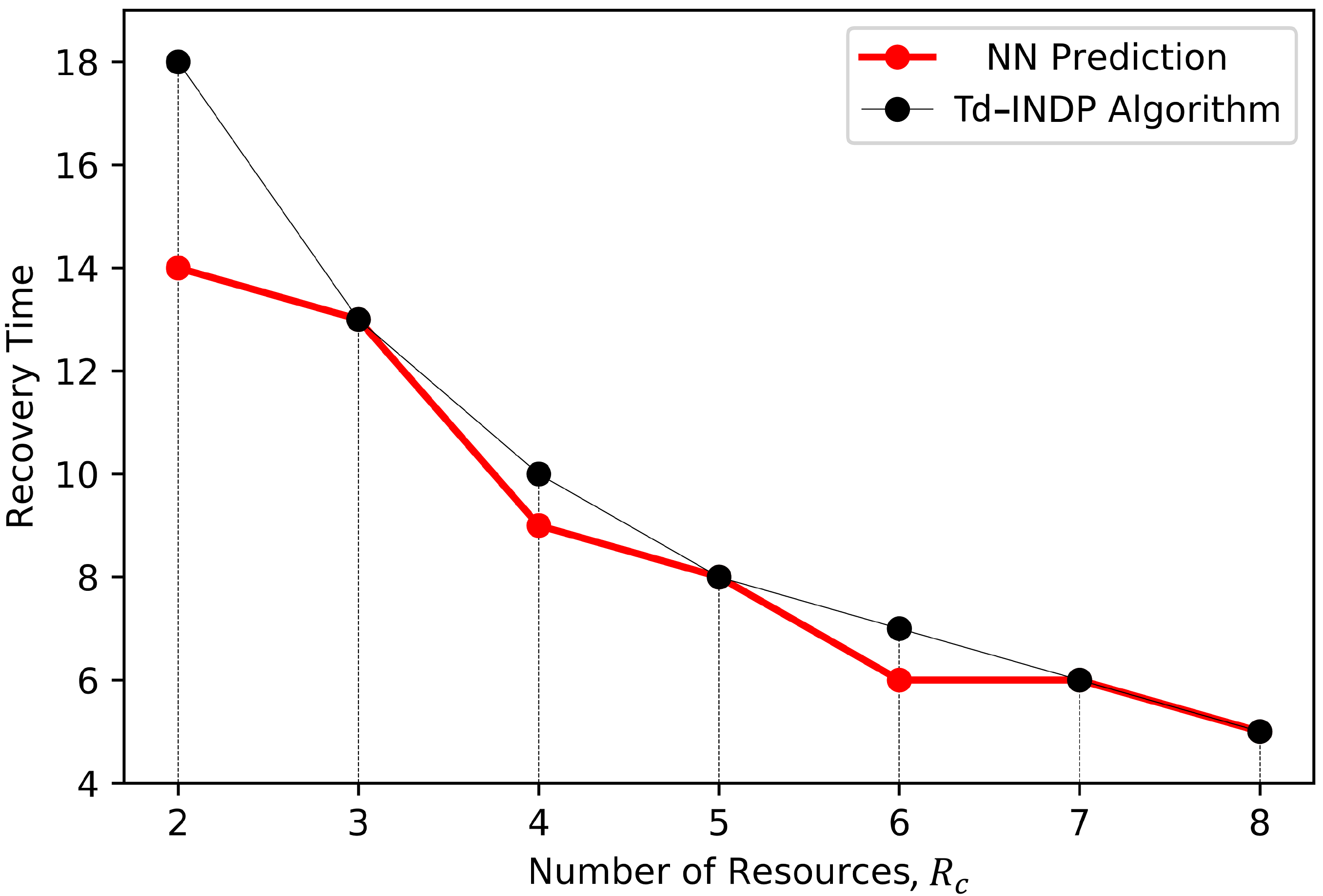}
        \caption{Network recovery time vs $R_c$, Shelby County, TN.}
        \label{fig:resource_allocation_results}
    \end{minipage}
\end{figure}




\subsection{Resource Allocation}
\label{subsec:rsc-all}

As mentioned in \Cref{subsec:NN-Based-Resource-Allocation}, in our problem setup, resource allocation comes into play in order to address a balance between a reasonable recovery time of the damaged portion of the network and the number of resources utilized to that end.
Based on our empirical results from td-INDP, we restrict $R_c\in\{2,3,\cdots,8\}$.
Given that the average number of damaged nodes is $11$ (for $m=6$) up to $57$ (for $m=9$), $R_c=10$, for instance, yields a speedy recovery (roughly two and six time-steps on average, for $m=6$ and $m=9$ respectively).
Such a short duration of recovery implies an abundance of resources, which makes resource allocation trivial and inefficient.
Therefore, we use a set of relatively small $R_c$ values in our analyses.\footnote{The reader is referred to Gonzalez et al. \cite{Gonzalez2016c} for more details on resource allocation settings of the simulator.}
Accordingly, we leverage the predictive power of the \ac{NN} model that we discussed in \Cref{subsec:NN-td-INDP} to train a separate model for each $R_c$.
The models are trained on every possible damage scenario and for different magnitudes of earthquake, $m$.
The outcome of all the trained models can be collected as a reference for the potential use of the decision-maker for a variety of scenarios.

\Cref{fig:damage_map} gives an example of a damage scenario with $m=7$ and depicts the pinned locations of damaged water, gas, and power nodes (respectively shown in blue, green, and red) in Shelby County, TN.
\Cref{fig:resource_allocation_results} demonstrates the time-step that all damaged nodes are recovered ($y$-axis) against $R_c$ ($x$-axis) and the optimal results from \ac{td-INDP} (black curve) are compared to the output of our trained model (red curve).
As indicated, one unit of increase in $R_c$, say from 3 to 4, can approximately save up to 3 time-steps.\footnote{Time-steps could possibly represent hours, days, or even weeks after the occurrence of a natural disaster. Hence, for a time-critical application, three time-steps can be a substantial saving.}
On the other hand, for applications in which the number of resources is limited, $R_c$ can be cut down with the cost of a longer recovery process.
Also, note that as $R_c$ gets larger, investing in more resources results in less improvement in the recovery time (compared to smaller $R_c$).

Finally, we have included the prediction of our trained model in the same plot for comparison.
The result shows a promising estimate of the underlying behavior of \ac{td-INDP}.
We acknowledge that such prediction may not be applicable for high-precision approximations; however, the computational complexity of using pre-trained models is significantly lower compared with solving a high-dimensional \ac{MIP}.
This is due to the fact that the running time to solve the \ac{MIP} of td-INDP can increase exponentially as the number of variables grows \cite{Gonzalez2016}.
For instance, it approximately takes 10 seconds for \ac{td-INDP} to find a single recovery plan on a Core i7-6700K 4.00 GHz quad-core processor for the Shelby County test case with $125$ infrastructure nodes.
However, the time to solve such \ac{MIP} can grow exponentially towards hours or even days of computations as the size of the system grows up to thousands or millions of nodes.


\subsection{Interdependency of the Networks}
\label{subsec:interpret_NN}

Besides the predictive power of the trained \ac{NN}s, we will use the features of the hidden layers to extract meaningful patterns and understand interdependencies in the infrastructure networks.
It is previously shown that the interdependency of the different components of the network is reflected in a recovery operator obtained from linear regression models \cite{gonzalez2017efficient}.
In a sense, our method generalizes these results by exploiting the structure of the trained \ac{NN} that incorporates the nonlinear structure of the problem.
To enhance visualization, we study the interpretability of a \ac{NN} model with one hidden node containing $n_h=4$ neurons.

\begin{figure}[t]
    \centering
    \includegraphics[width=0.6\columnwidth]{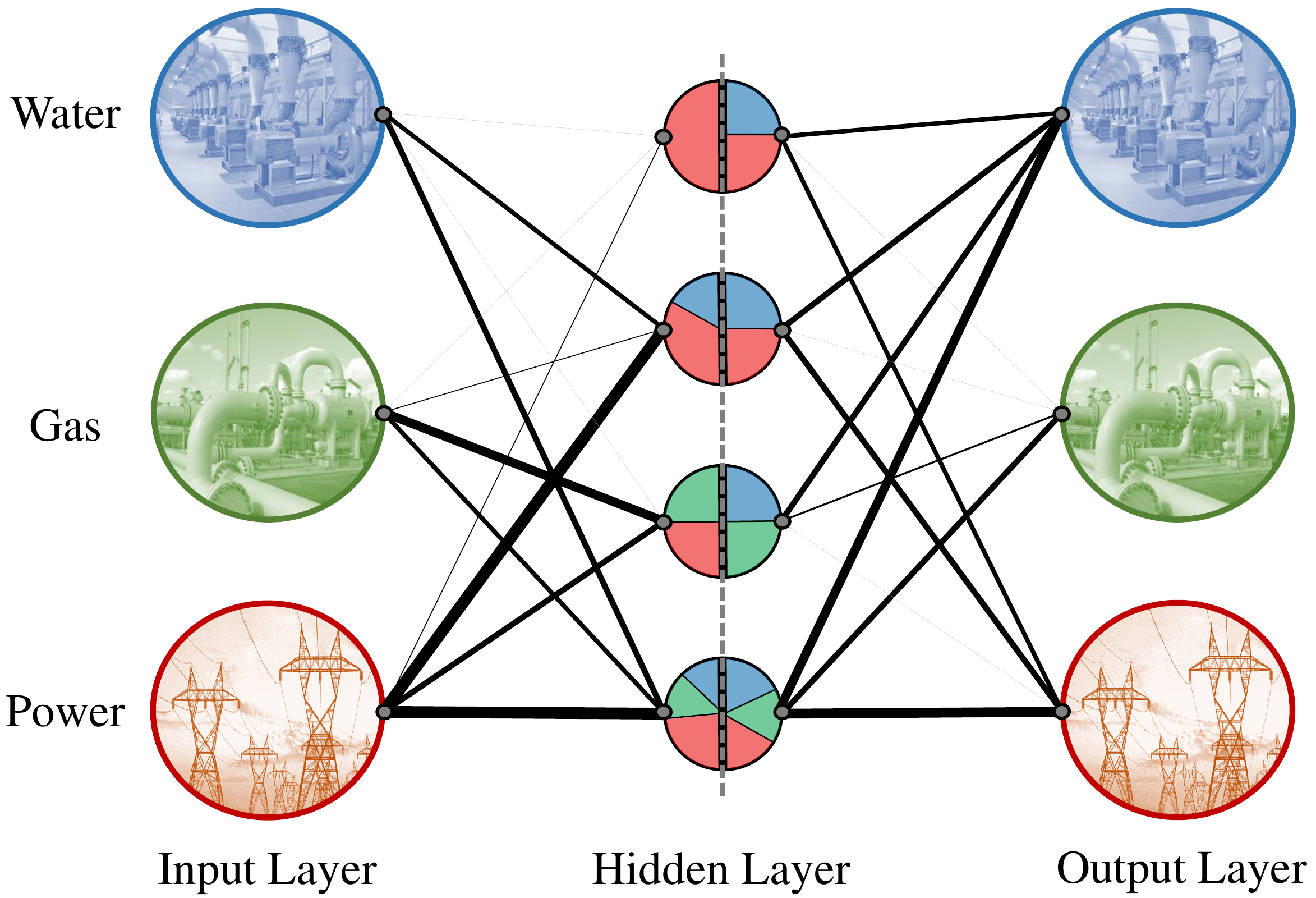}
    \caption{Visualization of the network interdependencies; thickness of the edges and color-coded regions on the hidden layer neurons imply qualitative influence.}
    \label{fig:structure}
\end{figure}

The results of our study are portrayed in \Cref{fig:structure}.
Dictated by their functionality, we divide the infrastructure nodes into three main divisions: water, gas, and power (with $49$, $16$, and $60$ nodes, respectively).
The edges connected to the hidden layer represent the trained weights during the learning phase.
These edges are resulted from the summation of all weights from each category; for example, to find the weight between the water nodes and any of the four hidden neurons, all $49$ corresponding water edges are aggregated.

We interpret nodes with larger corresponding weights as more influential nodes in the recovery process (denoted by the thickness of the edges).
Every neuron in the hidden layer captures some amount of information from the damaged sector of the network (input) and contributes to the recovery plan (output).
This is color-coded on each hidden neuron, which is, in turn, divided into left and right half-circles.
On each half, the amount of information is denoted by the area with the corresponding color.
A larger portion implies more transferred information.
In particular, the first two neurons capture the dependency of water network on the power system as the neurons are highly influenced by the power network in the input, and contribute to the prediction for both water and power networks almost equally.
Similarly, the third neuron captures information from the power network, and influences the recovery prediction of the water division in the output.
Note that the contribution of the gas network in the input and output for the last two neurons is almost symmetrical.
This coincides with the fact that the gas networks' recovery is physically independent of other networks, although the restoration of networks is coupled because of the shared resources.

\begin{figure}[t]
    \centering
    \begin{minipage}{.45\textwidth}
        \centering
        \centering
        \includegraphics[width=0.99\columnwidth]{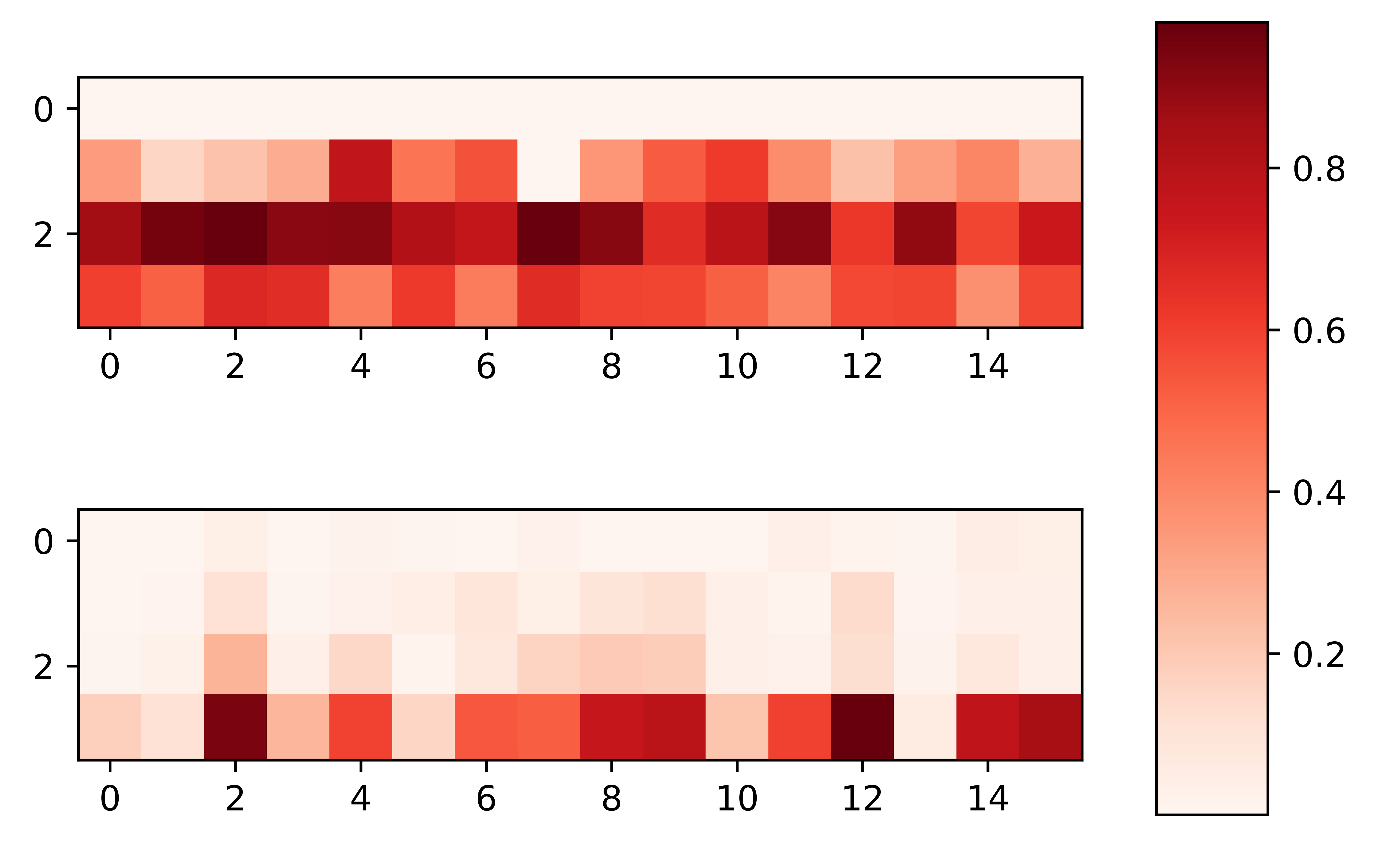}
        \caption{The heatmap of the weights between the four hidden neurons (y-axis) and the gas nodes (x-axis) in the input vector (upper plot) and the output (lower plot).}
        \label{fig:gasChess}
    \end{minipage}%
    \hspace{7mm}
    \begin{minipage}{.45\textwidth}
        \centering
    	\includegraphics[width=0.81\columnwidth]{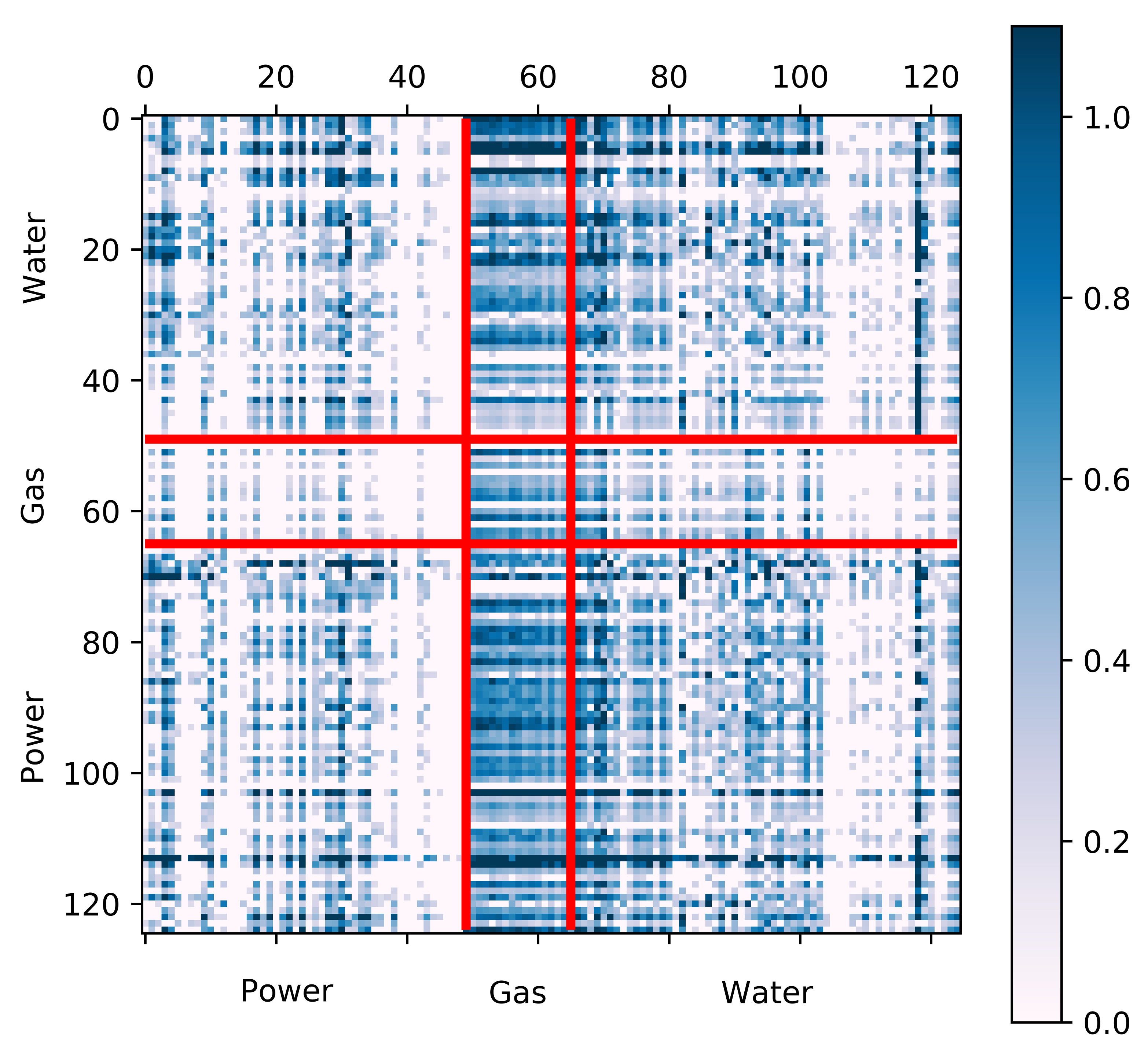}
    	\caption{Linear recovery operator obtained from trained \ac{NN} model. The rows determine the weights given to the input (binary vector with length $125$). The resulting vector from multiplying the operator by an input gives an approximation of the restoration strategy.}
    	\label{fig:Matrix}
    \end{minipage}
\end{figure}

On a node-level, we can represent the results above in the form of a heatmap.
\Cref{fig:gasChess} demonstrates a heatmap associated with the gas nodes on the overall restoration of \ac{NN}.
The upper plot encodes the weights between the four hidden neurons and the gas nodes in the input, and the lower heatmap shows the same for the output (darker colors imply larger weights).
The patterns verify the prior observations from \Cref{fig:structure}, for instance, the restoration of the gas nodes is mainly determined by the last two hidden layers (similar heatmaps of power and water networks are omitted due to limited space).


On the recovery-level, \ac{td-INDP} poses some prioritization in order to allocate resources.
\Cref{fig:Matrix} depicts such a pattern on a network-level.
The figure is an input-to-output $125\times 125$ recovery operator obtained from the \ac{NN} model described in \Cref{fig:structure}.
To build such an operator, we multiply the corresponding weights from input to each neuron in the hidden layer and then from there to the corresponding output.
Then, we aggregate the four outcomes from each hidden neuron to get a scalar weight from a desired input node to the output layer.
Hence, the linear operator encodes a mapping from a given damage scenario to a recovery sequence plan.

As the heatmap suggests, the recovery of the entire network is highly dependent on the recovery of the gas nodes (and not vice versa).
This is reasonable since the gas network is prioritized by \ac{td-INDP}, and hence, it highly affects the distribution of resources in early time-steps, although it is physically independent from other networks.
The water and power networks also display a moderate interdependency (i.e., recovery of water nodes heavily depends on gas nodes and is moderately influenced by the power division).
Finally, note that a similar approach was implemented in \cite{gonzalez2017efficient} by directly using a least-squares approach.
We emphasize that such results can be recovered from our model as \ac{NN} behaves similar to a linear operator in case we remove hidden layers.
However, the advantage of our machinery is that besides the more powerful prediction, we can also obtain much insight from the network interdependencies obtained from the hidden layer in \Cref{fig:structure}.



\section{Conclusions and Remarks}
\label{sec:conclusions}

In this paper, we study the resource allocation problem for infrastructure networks using machine learning techniques.
In particular, we train deep neural network models to circumvent the computational complexity that comes with the high-fidelity disaster recovery simulator, \ac{td-INDP}.
Employing the prediction power of such effective nonlinear models, we train multiple \ac{NN}s for different numbers of resources to find the most efficient plan for restoration in real-time.
We further explore our trained models to obtain meaningful insights about the underlying behavior of the simulator.
Finally, we produce a comprehensive dataset that can be used for further research or reproducing the results of this work which is publicly available along with implementation codes.

We acknowledge that our proposal is an initial step towards optimal restoration for infrastructure resilience and can be improved in many aspects.
One future direction is to systematically optimize over the number of resources, $R_c$, so as to address the trade-off between recovery time and the number of required resources.
We believe this can be tackled with both advanced optimization techniques and reinforcement learning methods for discrete control.
Another route to extend the results of this work is to implement similar machinery on networks with various topologies and configurations.
In that case, it would be essential to perform a sensitivity analysis on the parameters of \ac{td-INDP}.

\section{Acknowledgments}

The authors would like to thank Airlie Chapman, Andres Gonzalez, and Raissa D'Souza for insightful conversations and suggestions on the applications of network theory and prediction models on infrastructure networks.
The research of the authors has been
supported by NSF grants SES-1541025 and CMMI-1541033.


\bibliographystyle{ieeetr}
\bibliography{citations}



\section*{Appendix}
\label{sec:appendix}

\noindent \textbf{Source Code Instructions.}
For the implementation of this work, we have used \textit{python} 3.7.4 and \textit{PyTorch} 1.2 \cite{NEURIPS2019_9015}.
The reader is referred to the \href{https://github.com/siavash91/Td-INDP-NN}{GitHub repository} of this project for details on the dataset, code, and \ac{NN} settings.

\end{document}